\documentclass[aps,prl,superscriptaddress,floatfix,showpacs,notitlepage,twocolumn]{revtex4-1}
\usepackage[latin9]{inputenc}
\usepackage{color}
\usepackage{textcomp}
\usepackage{amsmath}
\usepackage{amssymb}
\usepackage{enumitem}
\usepackage{graphicx}
\usepackage{esint}
\usepackage{framed} 
\usepackage{xcolor}
\usepackage{hyperref}
\hypersetup{colorlinks,linkcolor=blue,citecolor=blue,filecolor=blue,urlcolor=blue}\makeatletter
\usepackage{amsfonts}
\usepackage{bbold}
\usepackage{natbib}

\pdfoutput=1
\newcommand{\ket}[1]{|#1 \rangle}
\newcommand{\bra}[1]{\langle#1 |}

\newcommand{\lr}[1]{\left( #1 \right)}

\newcommand{\mean}[1]{\langle #1 \rangle}

\newcommand{\mc}[1]{\mathcal{#1}}

\begin{document}

\title{Quantum Optimization for Maximum Independent Set Using Rydberg Atom Arrays}
\author{Hannes Pichler}
\thanks{These authors contributed equally to this work.}
\affiliation{ITAMP, Harvard-Smithsonian Center for Astrophysics, Cambridge, MA 02138, USA }
\affiliation{Department of Physics, Harvard University, Cambridge, MA 02138, USA }
\author{Sheng-Tao Wang}
\thanks{These authors contributed equally to this work.}
\affiliation{Department of Physics, Harvard University, Cambridge, MA 02138, USA }
\author{Leo Zhou}
\affiliation{Department of Physics, Harvard University, Cambridge, MA 02138, USA }
\author{Soonwon Choi}
\affiliation{Department of Physics, Harvard University, Cambridge, MA 02138, USA }
\affiliation{Department of Physics, University of California Berkeley, Berkeley, CA 94720, USA }
\author{Mikhail D. Lukin}
\affiliation{Department of Physics, Harvard University, Cambridge, MA 02138, USA }

\begin{abstract}
We describe and analyze an architecture for quantum optimization to solve maximum independent set (MIS) problems using neutral atom arrays trapped in optical tweezers.
Optimizing independent sets is one of the paradigmatic, NP-hard problems in computer science. 
Our approach 
is based on coherent manipulation of atom arrays via the excitation into Rydberg atomic states. 
Specifically, we show that solutions of MIS  problems can be efficiently encoded in the ground state of interacting atoms in 2D arrays by utilizing the Rydberg blockade mechanism.    
By studying the performance of leading classical algorithms, we identify parameter regimes, where computationally hard instances can be tested using near-term experimental systems. Practical implementations 
of both quantum annealing and variational quantum optimization algorithms beyond the adiabatic principle are discussed.  
\end{abstract}

\date{\today}

\maketitle

\textit{Introduction.---}
Quantum optimization is a paradigm to solve combinatorial optimization problems by utilizing controlled dynamics of quantum many-body systems \cite{Farhi:2001hya,Kadowaki:1998bm,Das:2008hz,{Albash:2018he}}. 
The key idea is 
to steer the dynamics of quantum systems such that their final states provide solutions to optimization problems~\cite{Das:2008hz}.
Such dynamics are often achieved either via the adiabatic principle \cite{Farhi:2001hya} in quantum annealing algorithms (QAA), or by employing more general, variational approaches,  as exemplified by quantum approximate optimization algorithms (QAOA) \cite{Farhi:2014wk,{Peruzzo:2014kc}}.
While these algorithms suggest an intriguing possibility to tackle computationally difficult problems beyond the capability of classical computers~\cite{Farhi:2016vm}, their heuristic nature makes it difficult to predict their practical performance and calls for  experimental tests.

Early experimental attempts involved directly annealing a number of superconducting systems coupled to a low temperature environment and generated a lively debate regarding the presence of quantum aspects associated with  this approach \cite{Johnson:2011gd,{Ronnow:2014fd},Boixo:2013ha,{Smolin:2013vp},{Wang:2013tg},{Shin:2014wf}}.
More recently, several experiments 
demonstrated programmable quantum dynamics of a large number of coupled qubits with high degree of quantum coherence \cite{Labuhn:2016dp,Bernien:2017bp,Lienhard:2018fm,GuardadoSanchez:2018jf,Zhang:2017et,{Monz:2016gz},{Barends:2016ju},{Otterbach:2017uh},{Neill:2018gi},{Houck:2012iqa},{Kaufman794},{Gross:2017do}}. 
While such quantum machines constitute promising platforms for testing quantum algorithms, the efficient implementation of quantum optimization in these systems remains an outstanding challenge.  

This Letter describes and analyzes an efficient method to realize quantum optimization by using neutral atoms excited into Rydberg states~\cite{Saffman:2010ky}.
Our approach harnesses recently demonstrated capabilities including (\textit{i}) deterministic positioning of individual neutral atoms in arrays with arbitrary arrangements~\cite{Endres:2016fka,Barredo:2016ea}, and $(ii)$ coherent manipulation of the internal states of these atoms, including excitation into Rydberg states \cite{Bernien:2017bp,Labuhn:2016dp,Lienhard:2018fm,GuardadoSanchez:2018jf}. 
We show that the combination of these capabilities allows us to efficiently  encode an important class of  NP-complete optimization problems. We describe explicit methods to implement both QAA and QAOA with numerical demonstrations and resource analysis, which reveals that near-term experiments can be used to test these algorithms for problem sizes beyond the capability of modern classical machines.
\begin{figure}[b]
\begin{center}
\includegraphics[width=0.5\textwidth]{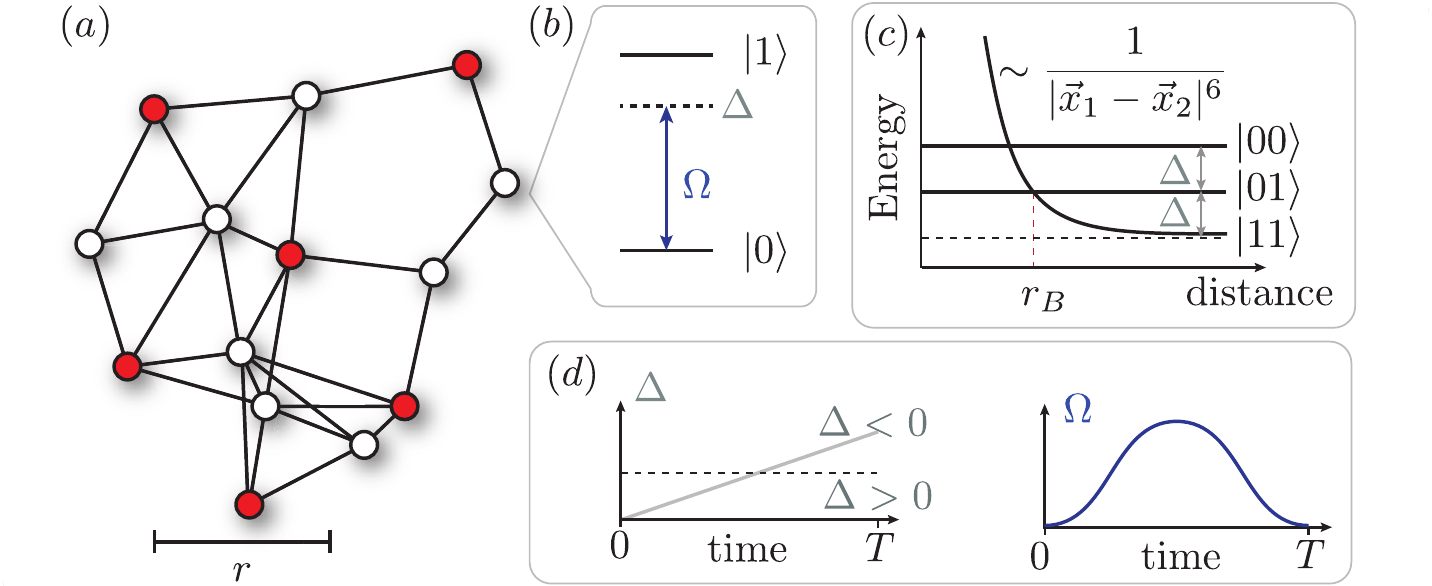}
\caption{Rydberg blockade mechanism and maximum independent set on unit disk graphs. (a) Example of a UD graph with the corresponding MIS indicated in red. (b) Vertices are represented by two-level atoms, that can be coherently driven with Rabi frequency $\Omega$ and detuning $\Delta$. (c) Interatomic interaction potentials in the limit of weak driving,  $\Omega\ll\Delta$ and $\Delta>0$. For atoms closer than $r_B$ it is energetically favorable for one of them to stay in $\ket{0}$. (d) Sweep profile for $\Omega(t)$ and $\Delta(t)$. }\label{Fig1PRL}
\end{center}
\end{figure}

\textit{Maximum independent set.---} We consider a specific, paradigmatic optimization problem called maximum independent set.
Given a graph $G$ with vertices $V$ and edges $E$, we define an independent set as a subset of vertices where no pair is connected by an edge [Fig.~\ref{Fig1PRL}(a)].
The computational task is to find the largest such set, called the maximum independent set (MIS).
Deciding whether the size of MIS is larger than a given integer $a$ for an arbitrary graph $G$ is a well-known NP-complete problem, and even approximating the size is NP-hard \cite{Garey:1979wr}.
This problem can be cast into a  spin model with Hamiltonian 
\begin{align}\label{eq:HMIS}
H_{P}=\sum_{v\in V} -\Delta \,n_v+\sum_{(v,w)\in E} U n_v n_w
\end{align}
where we assignd a spin-1/2 system with states $\ket{0}$ and $\ket{1}$ to each vertex, and defined $n_v=\ket{1}_v\bra{1}$.
When $U > \Delta >0$, $H_P$ energetically favors each spin to be in the state $\ket{1}$ unless a pair of spins are connected by an edge.
Hence, in the ground state of $H_P$ only the vertices in MIS are in state $\ket{1}$. 

A simple QAA can  be devised by adding a transverse field $H_D=\sum_{v}\Omega\sigma^x_v$ with $\sigma^x=\ket{0}\bra{1}+\ket{1}\bra{0}$, that induces quantum tunneling between different spin configurations [Fig.~\ref{Fig1PRL}(b)].
By slowly changing $\Omega$ and $\Delta$ in a time-dependent fashion, one can adiabatically connect the trivial ground state with all qubits in $\ket{0}$ ($\Delta<0,$ $\Omega=0$) to a final state encoding the MIS ($\Delta>0, \Omega=0$) and, in principle, solve the optimization problem [see Fig.~\ref{Fig1PRL}(d)].  Note that in general, such a procedure involves transitions across vanishing 
energy gaps as discussed in detail below. 

A particularly important class of graphs, which is intimately related to Rydberg physics, are so-called unit disk (UD) graphs. UD graphs are geometric graphs, where vertices are placed in the 2D plane and connected if their pairwise distance is less than a unit length, $r$ [see Fig.~\ref{Fig1PRL}(a)]. 
The MIS problem on UD graphs (UD-MIS) is still NP-complete~\cite{Clark:1990dz} and commonly appears in practical situations ranging from wireless network design~\cite{Hale:gn,Vahdatpour:2008hb} to map labelling \cite{Agarwal:1998fn} in various industry sectors~\cite{Wu:2006ju}~\footnote{We note that, unlike generic MIS, UD-MIS allows polynomial-time algorithms to approximate the solution \cite{Matsui:1998dy,Nandy:2017ij}}. 

\textit{Rydberg implementation.---} We now turn to a physical system that allows us to tackle this problem using state-of-the-art experimental technology. Recently, it has been demonstrated that a large number of atoms can be individually and deterministically placed in a 2D plane using optical tweezers \cite{Labuhn:2016dp,Lienhard:2018fm}.
Each atom, realizes a qubit, $v$, with an internal ground state, $\ket{0}$, and a highly excited, long-lived Rydberg state, $\ket{1}$, which can be coherently manipulated by external laser fields. The Hamiltonian governing the evolution of such a system is 
\begin{align}\label{eq:HRYD}
H_{\rm Ryd}\!=\!\sum_{v}\! \left( \Omega_v \sigma_v^x-\Delta_v n_v \right) +\!\sum_{v< w}\! V(|\vec{x}_v\!-\!\vec{x}_w|)\,n_v n_w,
\end{align}
where $\Omega_v$ and $\Delta_v$ are the Rabi frequency and laser detuning at the position of atom, $\vec{x}_v$. 
While individual manipulation is feasible, we focus on a global, homogeneous driving laser, i.e., $\Omega_v=\Omega$ and $\Delta_v=\Delta$, and isotropic Rydberg interactions decaying as $V(x)=C/x^6$.
The interactions effectively prevent two atoms from being simultaneously in state $\ket{1}$ if they are within the Rydberg blockade radius $r_B=\lr{C/\sqrt{(2\Omega)^2 + \Delta^2}}^{1/6}$ [see Fig.~\ref{Fig1PRL}(c)],
providing the natural connection between the Rydberg Hamiltonian $H_{\rm Ryd}$ and UD-MIS problems.
For the special case of 1D arrays, this (Fig.~\ref{Fig1PRL}(d)) setting has been realized in recent quantum simulation experiments \cite{Bernien:2017bp}.
\begin{figure}
\begin{center}
\includegraphics[width=0.5\textwidth]{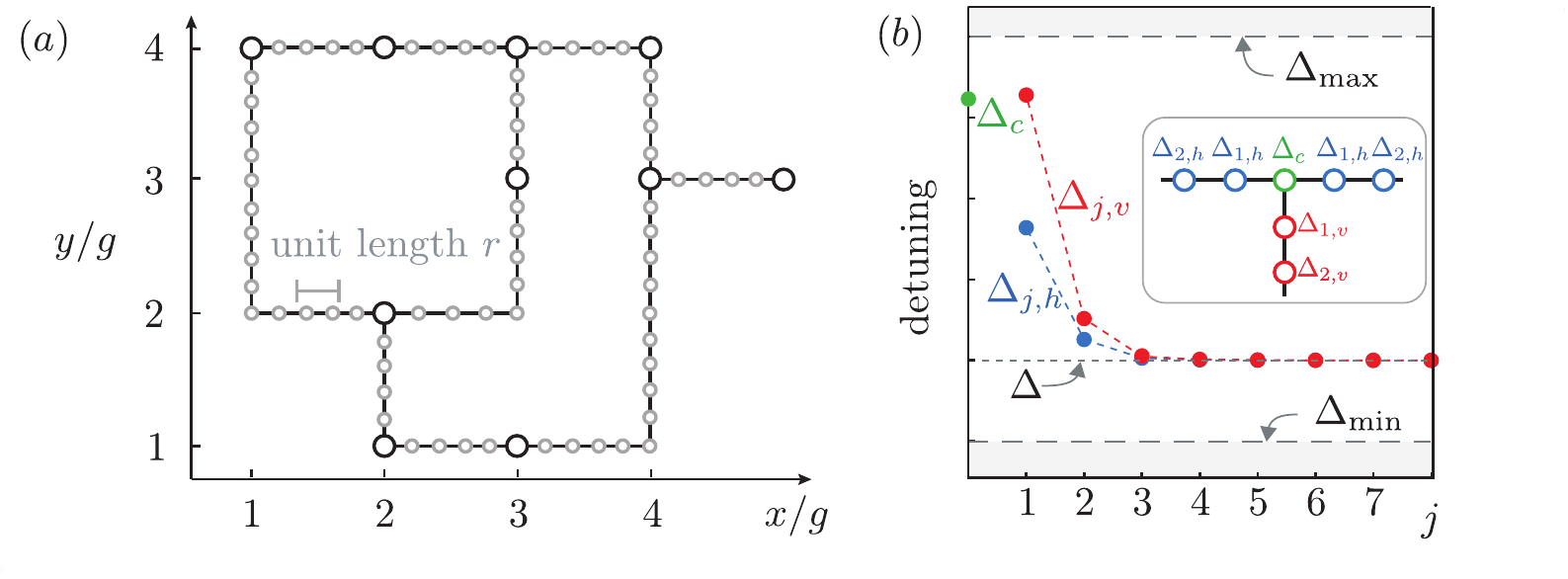}
\caption{NP-hardness of the Rydberg Hamiltonian $H_{\rm Ryd}$.  (a) A planar grid graph with maximum degree 3, $\mc{G}$ (big circles), can be embedded on a grid and transformed into a unit disk graph, $G$, using ancillary vertices (small circles) along the edges of $\mc{G}$. 
(b) A particular choice of detunings close to a junction compensates for the effects of long range dipolar interactions. Here, $\Delta_{j,v(h)}$ is the detuning of an atom at distance $j$ from the central atom, with detuning $\Delta_c$, along the vertical (horizontal) direction (see inset). 
}\label{Fig2PRL}
\end{center}
\end{figure}

\textit{NP-hardness.---}
One can rigorously show that the Rydberg Hamiltonian \eqref{eq:HRYD} can exactly encode NP-complete problems, even in the presence of the long-range interaction tails.
To prove this, we provide an explicit reduction, which transforms each instance of a NP-complete problem into an arrangement of atoms, such that the ground state of \eqref{eq:HRYD} encodes the solution \cite{paper_proof}.
The key idea is to generalize the reduction from MIS on planar graphs with maximum degree 3 to UD-MIS \cite{Clark:1990dz,doi:10.1137/0132071}. 
A graph of the former type, $\mc{G}$, 
can be efficiently embedded on a grid (with grid unit $g$), and transformed to a UD graph, $G$, by introducing an even number of ancillary vertices
along each edge 
(see Fig.~\ref{Fig2PRL}(a) and \cite{paper_proof}). The UD radius $r\simeq\frac{g}{2k+1}$ is determined by a parameter $k$, proportional to the linear density of ancillary vertices. We choose detunings $\Delta_v$ in a range $\Delta_{\rm min}\equiv 0.51C/r^6<\Delta_v< \Delta_{\rm max}\equiv C/r^6$, and a sufficiently large $k$ \cite{paper_proof}.
Then, in the ground state of $H_{\rm Ryd}$ (at $\Omega_v=0$), each array of ancillary atoms is in an ordered configuration, alternating between $\ket{0}$ and $\ket{1}$ (with at most one domain wall). Moreover, we can show that the tail of the Rydberg interactions is only relevant in the vicinity of corners and junctions, where arrays of ancillary atoms meet under an angle (Fig.~\ref{Fig2PRL})\cite{paper_proof}.
Crucially, one can adjust the detuning pattern around these structures to compensate for the effect of interaction tails (see Fig.~\ref{Fig2PRL}(b) for an explicit example of such a detuning pattern). This ensures that the ground state of $H_{\rm Ryd}$ \eqref{eq:HRYD} for $\Omega_v=0$, coincides with the one of the MIS Hamiltonian \eqref{eq:HMIS} on $G$. We stress that this procedure does not require any a priori knowledge of the MIS \cite{paper_proof}. 
Since the size of the MIS on $G$ is identical to that of the MIS on $\mc{G}$ up to half of the number of ancillary vertices, it is NP-hard to find the ground state of the Rydberg Hamiltonian \eqref{eq:HRYD} \cite{paper_proof}.

\textit{Generalizations. --- }
One can also employ various optical techniques to enlarge the class of problems that can be addressed in these systems. One powerful approach involves qubits encoded in hyperfine ground states, and selective excitation into various kinds of Rydberg states \cite{{Glaetzle:2014cg},Glaetzle:2015iu,{Zeiher:2016td}}.  
In particular, one can utilize strong and long-range dipolar interactions between Rydberg states with different parities $S$ and $P$ \cite{{Thompson:2017gk},Saffman:2010ky}, to efficiently realize rotations of a single qubit, controlled by its multiple neighbors. For example, one can excite all neighbors of a given (central) spin selectively from the state $\ket{1}$ to an $S$-state. Subsequently, two-step transition between the two hyperfine states of the central spin, via the Rydberg $P$-state, implements a multi-qubit controlled rotation \cite{SM}.  Combined with Trotterized time evolution, these techniques provide means to go beyond UD graphs, and implement various quantum optimization algorithms for MIS problems on arbitrary graphs \cite{SM}\footnote{This is complementary to recent QAA proposals that enlarge the number of qubits to achieve arbitrary connectivity \cite{Lechner:2015iy,Glaetzle:2017hh}}.

\begin{figure}[t]
\begin{center}
\includegraphics[width=\linewidth]{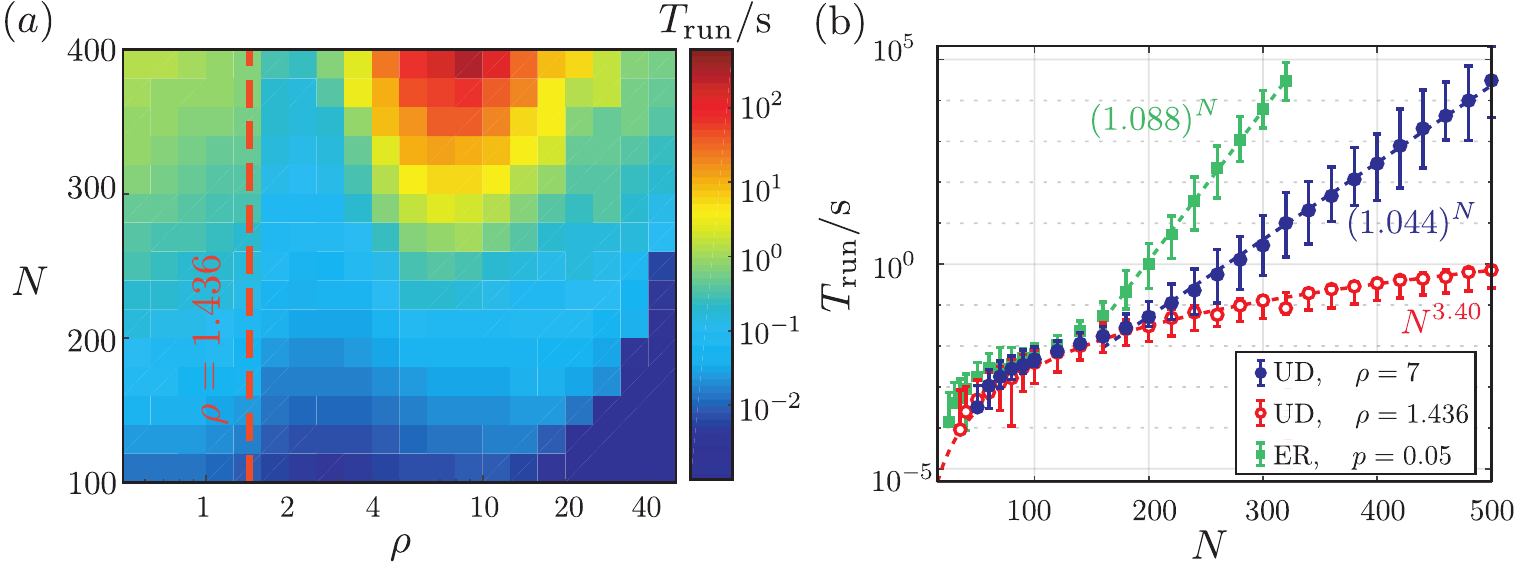}
\caption{Performance of a classical branch and bound algorithm \cite{Li:2017ik} on random UD-MIS. (a) Median runtime $T_{\rm run}$ to find the MIS in CPU time. Statistics are obtained from 50 graphs per data point. The dashed line indicates the percolation threshold. (b) At the percolation threshold $\rho$\,=\,$\rho_c$\,$\approx$\,1.436 (red) $T_{\rm run}$ scales polynomially with system size. For $\rho=7$ (blue), there is a clear exponential scaling with problem size for $N\gtrsim 150$. For comparison we also plot the performance on Erd\"os-Renyi graphs with edge probability $p=0.05$ (green),  showing a similar exponential runtime.
Error-bars are 5th and 95th percentile values among 100 graphs.}
\label{Fig:Classical}
\end{center}
\end{figure}

\textit{Optimization on random UD graphs ---} 
To identify regimes and system sizes where quantum algorithms could prove useful for UD-MIS, we now provide numerical simulations of both classical and quantum algorithms. We consider MIS on random UD graphs, constructed by placing $N$ vertices randomly with a density $\rho$ in a 2D box of size $L \times L$, where $L = \sqrt{N/\rho}$ and UD radius $r = 1$.
The hardness of this problem depends on the vertex density: If the density is below the percolation threshold $\rho_c\approx1.436$ \cite{Mertens:2012fr}, the graph decomposes into disconnected, finite clusters, 
allowing for efficient polynomial time algorithms. 
In the opposite limit, i.e.~if the density is very large, $\rho\rightarrow\infty$ (and $N\propto \rho$), the problem becomes essentially the closest packing of disks in the (continuous) 2D plane, with the known best packing ratio of $\pi/\lr{2\sqrt{3}}$.

In Fig.~\ref{Fig:Classical}(a), we analyze the performance of classical optimization algorithms, focussing on a  recently introduced, state-of-the-art branch and bound algorithm~\cite{Li:2017ik}. While for $\rho<\rho_c$ one finds a polynomial scaling [Fig.~\ref{Fig:Classical}(b)], at an intermediate density $\rho\sim 7$, the runtime exhibits a clear exponential dependence with system sizes exceeding $N\sim 150$ vertices, both on worst cases and typical instances [Fig.~\ref{Fig:Classical}(c)]. Note that at this density, the classical algorithm often could not find a solution for $N\gtrsim 440$ within 24 hours of CPU time on a single node of the Odyssey computing cluster \cite{Odyssey}.
For more general, non-UD graphs, this limitation of the classical algorithm is observed at $N \gtrsim 320$ \cite{DIMACS}. 

\begin{figure}[t]
\begin{center}
\includegraphics[width=1\linewidth]{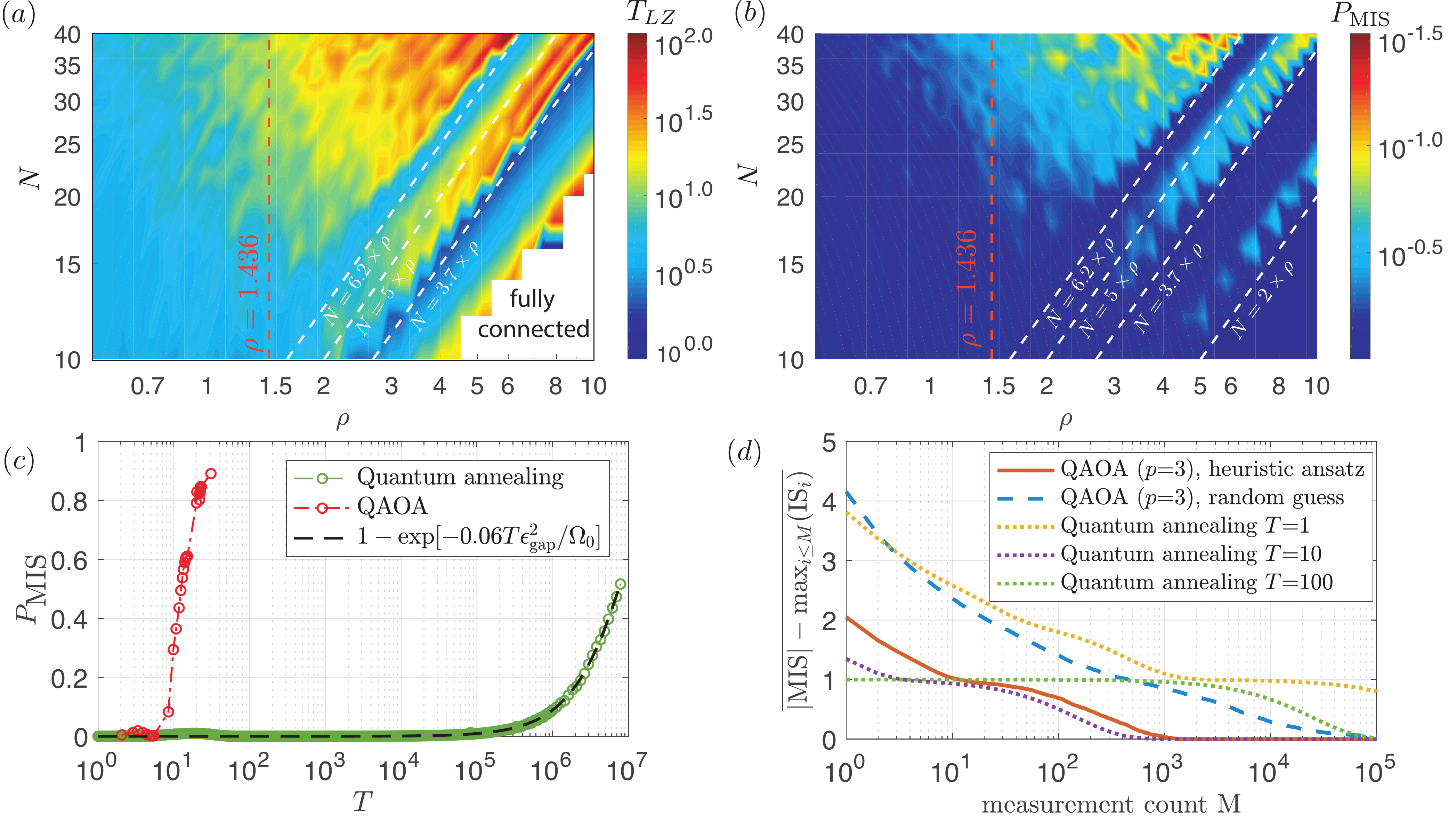}
\caption{Performance of quantum algorithms for MIS on random UD graphs. 
(a) Hardness diagram for quantum adiabatic algorithm in terms of adiabatic time scale $T_{\text{LZ}}$. White dashed lines correspond to optimal disk packing \cite{Musin:2015dz}. The fully connected region has trivially $|\text{MIS}| = 1$. (b) Success probability for non-adiabatic QAA with $T$\,=\,$10/\Omega_{0}$.
(c) Comparison of QAA with sweep \eqref{Ramp} and QAOA with optimized variational parameters of depth $p$ (different points correspond to increasing circuit depth up to $p = 50$), for a particular graph $G_0$ with $N=32$, $\rho = 2.4$ and $|\text{MIS}| = 8$, $\epsilon_{\rm gap} \approx 0.0012\Omega_0$ \cite{SM}.
(d) Average difference between $|\textrm{MIS}|$ and the larges independent set found after $M$ measurements for simulated QAOA and QAA on $G_0$. Dashed blue (solid red) line corresponds to QAOA with random initial values (heuristic ansatz) of $t_k,\phi_k,$.  
Dotted lines correspond to QAA followed by simulated measurements in the computational basis. For all QAA simulations we use sweeps \eqref{Ramp} with $\Delta_0/\Omega_0$\,=\,$6$.}
\label{Fig:QAOA}
\end{center}
\end{figure}

We now turn to the analysis of \textit{quantum} algorithms. In Fig.~\ref{Fig:QAOA}(a) we numerically simulate the adiabatic quantum annealing algorithm, with the MIS annealing Hamiltonian $H_D+H_P$ described above, sweeping the parameters as 
\begin{align}\label{Ramp}
\Delta(t)=\Delta_0(2t/T-1),\quad \Omega(t)=\Omega_0\sin^2(\pi t/T).
\end{align}
We take the limit $U/\Omega_0\rightarrow \infty$, where the dynamics is restricted to the independent set subspace, allowing us to numerically simulate system sizes up to $N\sim40$ qubits. 
We extract the Landau-Zener time scale, $T_{\text{LZ}}$, required for adiabaticity by fitting numerical results to the expected long time behavior of the ground state probability $P_{\rm MIS} =1 - e^{a-T/T_{\text{LZ}}}$ \cite{SM}. Again, we find a clear transition at the percolation threshold. At high density the size of the MIS should be determined by $L=\sqrt{N/\rho}$ owing to the relation to the optimum packing problem \cite{Musin:2015dz}. Indeed we find that this fact is also reflected in $T_{\rm LZ}$ as clearly visible by the stripe pattern in Fig.~\ref{Fig:QAOA}(a). The exponential Hilbert space size limits numerical simulations, inhibiting a conclusive comparison with the scaling of the classical algorithm above. In the experiments,  finite coherence times will further limit the performance of the adiabatic algorithms with exponential scaling of the minimum energy gap, $\epsilon_{\rm gap}$ \cite{Altshuler:2010cta,Laumann:2015gn}.

A number of approaches have been proposed to overcome this limitation.
These include heuristics to open up the gap \cite{Dickson:2011hy},  the use of diabatic transitions in QAA \cite{{Crosson:2014tg}, {Albash:2018he}}, and variational quantum algorithms such as QAOA \cite{Farhi:2014wk, MaxCutpaper} studied below.
In Fig.~\ref{Fig:QAOA}(b), we  show the probability to find the MIS, $P_{\rm MIS}$ for simulated non-adiabatic QAA \cite{{Crosson:2014tg}, {Albash:2018he}}. Specifically, we use a single sweep with profile in Eq.~\eqref{Ramp} with a short annealing time $T = 10/\Omega_{0}\ll T_{\rm LZ}$. 
Remarkably, the results show that substantial overlap with the MIS state can be obtained with evolution time $T\ll T_{\rm LZ}$, while the success probability resembles qualitatively the adiabatic hardness diagram. These results are consistent with previous observations that non-adiabatic annealing can be advantageous, especially when the minimum spectral gap is small \cite{{Crosson:2014tg}}. 

More recently, variational approaches, such as QAOA have been proposed for quantum optimization. 
In our case, a QAOA (of level $p$) consists of applying a sequence of $p$ resonant pulses of varying duration, $t_k$, and phase, $\phi_k$ on an initial state. This generates a variational wavefunction 
$\ket{\psi(\{t_k,\phi_k\})}$\,=\,$\prod_{k=1}^{p}\mc{U}_k\ket{0}^{\otimes N}$, where $\mc{U}_k$\,=\,$e^{-iH_k t_k}$ and $H_k=\sum_{v\in V}\Omega_0(e^{i\phi_k}\ket{0}_v\bra{1}+\textrm{h.c.})+\sum_{(v,w)\in E} U n_v n_w$, with $U/\Omega_0\rightarrow\infty$ \cite{SM}. 
The variational parameters $t_k$ and $\phi_k$ are optimized in a classical feedback loop from measurements of $H_P$ in this quantum state \cite{SM,Farhi:2014wk}. 
The performance of QAOA thus also depends on this classical optimization routine. In \cite{MaxCutpaper} we provide an in-depth analysis of various optimization strategies for large-$p$ QAOA, including a heuristic ansatz for $t_k$ and $\phi_k$. 
In Fig.~\ref{Fig:QAOA}(c), we compare optimized QAOA with quantum annealing \eqref{Ramp}, on a particular graph instance, $G_0$, (with $N$\,=\,$32$ and independent set space dimension $\rm{dim}_{IS}$\,=\,$17734$) for which the adiabatic timescale is $\sim$\,$10^6/\Omega_{0}$. We find that QAOA achieves $\mc{O}(1)$ success probability at depth $p\lesssim 40$ requiring significantly shorter evolution times, $T$\,$\equiv$\,$\sum_{k} t_k \sim 10/\Omega_0$\,$\ll$\,$T_{\rm LZ}$, thus demonstrating the existence of exceptionally good variational states. We expect this to underlie the performance of QAOA on large systems. For numerically accessible system sizes, however, the fluctuations associated with measurement projection noise typically imply that one  finds the MIS size during the classical optimization routine, even at smaller $p$.  
In Fig.~\ref{Fig:QAOA}(d), we illustrate this, with a full Monte Carlo simulation of QAOA on $G_0$, including sampling from projective measurements  \cite{SM}. The size of the largest measured independent set is plotted as a function of the number of measurements (averaged over Monte Carlo trajectories). With a properly choosen heuristic ansatz for $t_k$ and $\phi_k$ \cite{SM,MaxCutpaper}, QAOA (at $p=3$) finds the MIS already after $\sim 10^2$-$10^3$ measurements. This is comparable to the number of measurements required for accurately estimating $\mean{H_P}$, as part of the classical optimization routine.
This performance can be compared to repeated runs of QAA with sweeps \eqref{Ramp} for non-adiabatic annealing times $T=1/\Omega_0,10/\Omega_0,100/\Omega_0$. We observe that for this small system size, QAOA performs similarly as non-adiabatic QAA at the annealing time $T\sim10/\Omega_0$ [Fig.~\ref{Fig:QAOA}(d)]. We emphasize that for a more conclusive assessment of these algorithms, it is necessary to  test them using coherent experimental systems with more than 50 qubits. 

\textit{Experimental considerations. ---} Recent experiments with individually trapped Rydberg atoms have demonstrated deterministic placement of up to $\sim 50$ atoms in 1D \cite{Bernien:2017bp}, 2D \cite{Barredo:2016ea} and even 3D \cite{Barredo:2017uq}. Moreover, highly coherent dynamics based on the Rydberg blockade interactions has been achieved with high entanglement fidelities \cite{Levine:2018uk}. In particular, 1D experiments showed that (quasi) adiabatic sweeps for atoms arranged in regular lattices, find the corresponding MIS solution with a success probability reaching $\sim10\%$ \cite{Bernien:2017bp,QKZMpaper}. 
Based on realistic estimates of Rabi frequencies, $\Omega\sim 2\pi\times 10\,$-$100\,$MHz, and single qubit coherence times, $\tau\sim 200\mu$s (limited by Rydberg state decay \cite{Dephasing_footnote}), one can achieve a fully coherent QAOA circuit of volume $p N\simeq \Omega \tau > 10^4$ in near term experiments. In particular, extending these high-fidelity control techniques to 2D systems should enable running large $p$ QAOA for computationally hard problems of size $N\sim 10^2$-$10^3$. 
We emphasize that these are very conservative estimates, requiring less than one error in the entire array during the entire coherent evolution. Understanding the sensitivity of quantum optimizers to imperfections, exploring approaches to detect and suppress errors, and testing the scaling of quantum algorithms are among the central tasks that can be addressed within this platform. 

\textit{Conclusion and outlook. ---} We described a new approach for a hardware efficient implementation of quantum optimization based on neutral, trapped atoms, interacting via Rydberg states. Our proposal employs currently available technologies, encodes optimization problems directly without overhead, and is thus a promising candidate to test for a potential quantum speedup of different quantum optimization algorithms in near term devices. 
Our approach allows a direct experimental probe of the relation between quantum phase transitions in disordered spin systems and the complexity of combinatorial problems.   Diverse potential applications in areas ranging from network design \cite{Hale:gn,Vahdatpour:2008hb} to machine learning  \cite{Farhi:2018wv,Otterbach:2017uh} can be explored.  Finally, the explicit connection between the MIS problem and Rydberg blockade physics established in this work may be further developed and harnessed to investigate quantum many-body physics, such as quantum dimer models \cite{Moessner:2011kp,{Laumann:2012hu}}, Fibonacci anyons \cite{Feiguin:2007bo}, and quantum scars \cite{Turner:2018iz,{Ho:2018tv},{Khemani:ur}}. 

\textit{Acknowledgements. ---} We thank H.~Bernien, E.~Farhi, A.~Harrow, A.~Keesling, H.~Levine,  A.~Omran and P. Zoller for useful discussions. This work was supported through the National Science Foundation (NSF), the Center for Ultracold Atoms, the Air Force Office of Scientific Research via the MURI, the Vannevar Bush Faculty Fellowship and DOE. H.P. is supported by the NSF through a grant for the Institute for Theoretical Atomic, Molecular, and Optical Physics at Harvard University and the Smithsonian Astrophysical Observatory. S.C.~acknowledges support from the Miller Institute for Basic Research in Science.  Some of the computations
in this paper were performed on the Odyssey cluster supported by the FAS Division of Science, Research Computing Group at Harvard University.

\bibliography{HannesBib,additional_refs}	

\begin{thebibliography}{68}
\bibitem [{Dep()}]{Dephasing_footnote}%
  \BibitemOpen
  \href@noop {} {}\bibinfo {note} {Dephasing due to atomic motion, can result
  in additional errors. However, such effects can be circumvented by using spin
  echo or Raman sideband cooling.}\BibitemShut {Stop}%
\end{thebibliography}
\end{document}


\title{Supplemental Material: \\Quantum Optimization for Maximum Independent Set Using Rydberg Atom Arrays}

\author{Hannes Pichler}
\affiliation{ITAMP, Harvard-Smithsonian Center for Astrophysics, Cambridge, MA 02138, USA }
\affiliation{Department of Physics, Harvard University, Cambridge, MA 02138, USA }
\author{Sheng-Tao Wang}
\affiliation{Department of Physics, Harvard University, Cambridge, MA 02138, USA }
\author{Leo Zhou}
\affiliation{Department of Physics, Harvard University, Cambridge, MA 02138, USA }
\author{Soonwon Choi}
\affiliation{Department of Physics, Harvard University, Cambridge, MA 02138, USA }
\author{Mikhail D. Lukin}
\affiliation{Department of Physics, Harvard University, Cambridge, MA 02138, USA }

\begin{abstract}
In this supplemental material, we provide details on the numerical simulations of the quantum algorithms analyzed in the main text. We also elaborate on a potential generalization of the implementation discussed in the main text that allows one to address MIS problems on more general graphs than UD graphs with individually trapped Rydberg atoms. 
\end{abstract}

\date{\today}

\maketitle

\section{Quantum optimization on Maximum Independent Set}

In this section, we provide details on the numerical analysis of the various quantum optimization algorithms presented in the main text. 

\begin{figure}[b]
\begin{center}
\includegraphics[width=0.4\linewidth]{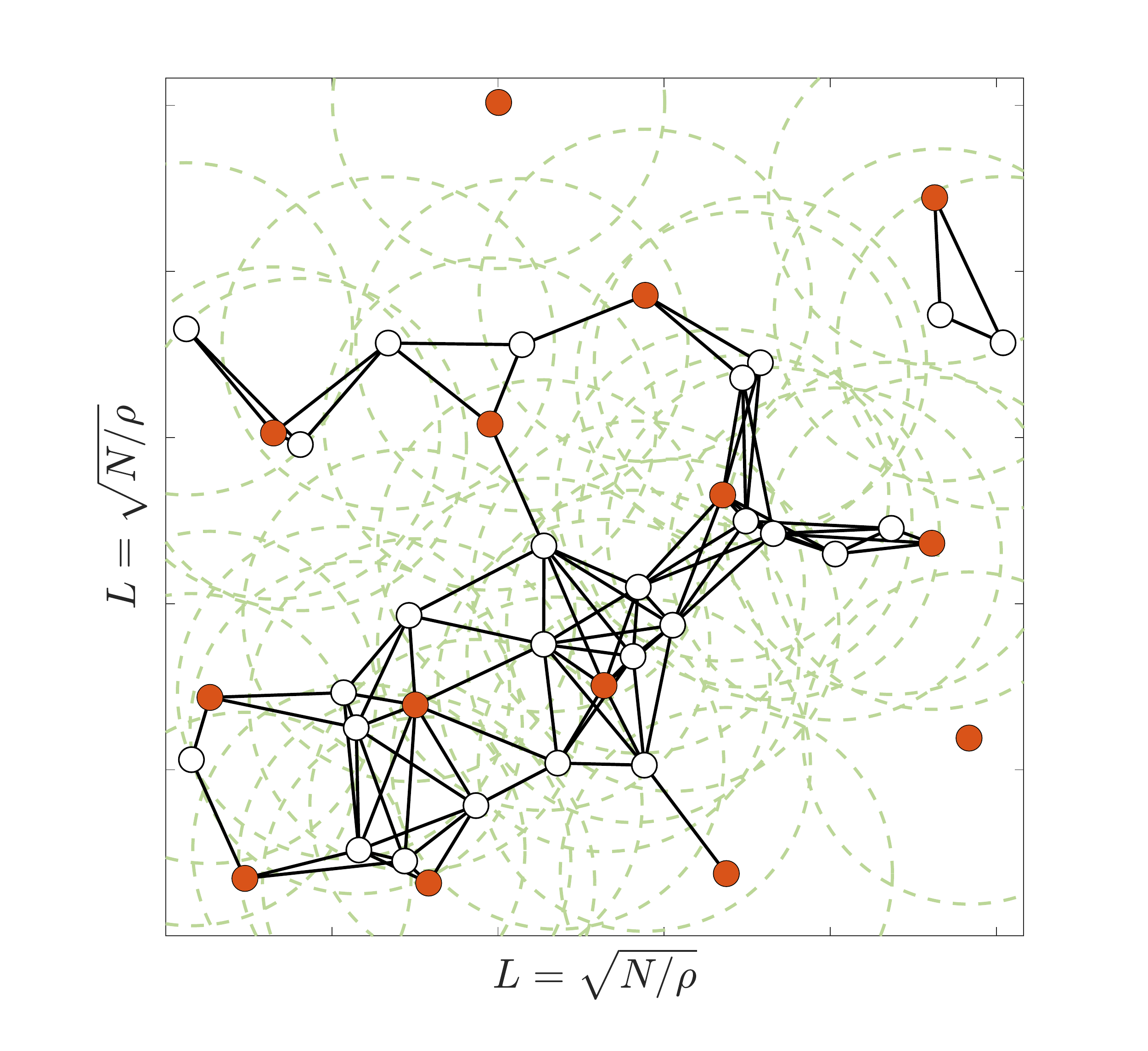}
\caption{An example of a random unit disk graph with $N = 40, \rho =1.5, |\text{MIS}| = 14$, and $93$ edges. The unit distance is set to be $r = 1$, and the box length is $L = \sqrt{N/\rho}$.}
\label{Fig:rUDGraph}
\end{center}
\end{figure}

As discussed in the main text, we focus on the maximum independent set problem on random unit disk (UD) graphs. We parametrize random UD graphs by two parmeters, the number of vertices $N$ and the 2D vertex density $\rho$. The unit distance is taken to be $r = 1$, and the vertices are put into a box of $L \times L$, where $L = \sqrt{N/\rho}$ (see Fig.~\ref{Fig:rUDGraph} for an example). For a random UD graph with density $\rho$, the average vertex degree is approximately $\pi \rho$. To minimize the finite-size boundary effect, we use periodic boundary conditions for UD graphs in all our numerical simulations.

\subsection{Quantum annealing for random UD-MIS}
As discussed in the main text, a quantum annealing algorithm (QAA) for MIS can be performed using the following Hamiltonian 
\begin{align} \label{Eq:QA}
H_{\text{QA}} (t)=\sum_{v\in V} \bigg(-\Delta(t) \, n_v+\Omega(t)  \sigma_v^x\bigg)+\sum_{(u,w)\in E} U n_u  n_w.
\end{align}
The QAA can be designed by first initializing all qubits  in $\ket{0}$ at time $t=0$, which is the ground state of $ H_{\text{QA}}(t=0)$ when $\Delta(t=0)<0$ and $\Omega(t=0)=0$ (with $U>0$). We then change the parameters, by first turning on $\Omega(t)$ to a non-zero value, sweeping $\Delta(t)$ to a positive value, and finally turning off $\Omega(t)$ again. The annealing protocol we consider throughout this work is specified by 
\begin{equation} \label{Eq:AdiabaticPath}
\Delta(t) = (2s-1)\Delta_0, \quad \Omega(t) = \Omega_{0} \sin^{2}(\pi s) \quad \text{with}  \quad s = t/T. 
\end{equation}
If the time evolution is sufficiently slow, then by the adiabatic theorem, the system follows the instantaneous ground state, ending up in the solution to the MIS problem. We take $\Omega_{0} = 1$ to be the unit of energy, and fix $\Delta_{0}/\Omega_{0} = 6$, which empirically seems to be a good ratio to minimize nonadiabatic transitions.

\begin{figure}[t]
\begin{center}
\includegraphics[width=0.5\linewidth]{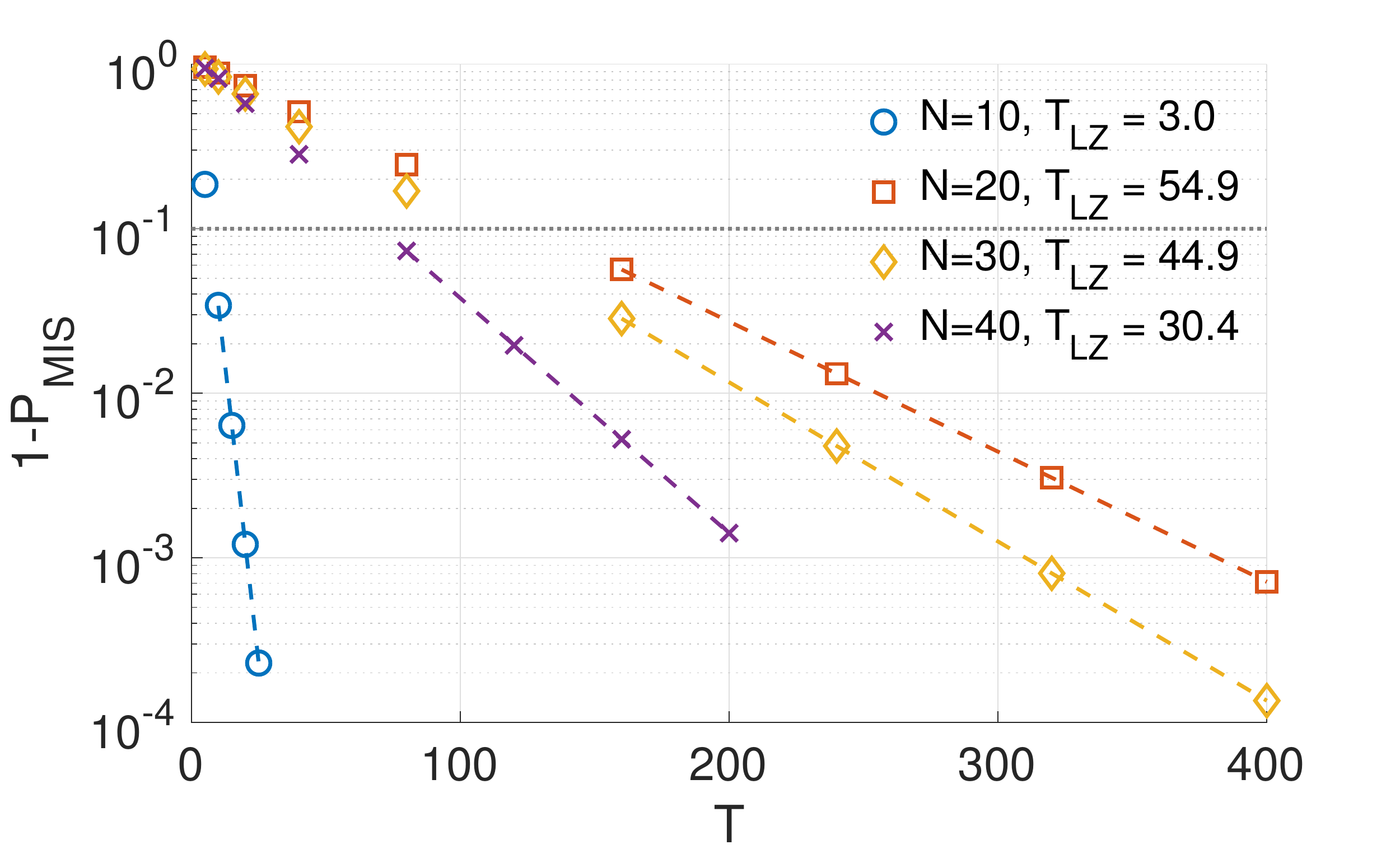} 
\caption{The Landau-Zener fitting to $1-P_{\text{MIS}} = e^{a-T/T_{\text{LZ}}}$ to extract the adiabatic time scale $T_{\text{LZ}}$. Here, four random unit disk graphs with $N = 10,20,30,40$ are shown. For each instance, we find the first $T$ iteratively such that $P_\text{MIS}(T) > 0.9$, denoted as $T^{*}$. The fitting is then performed on four points $T^{*}, 1.5T^{*},2T^{*},2.5T^{*}$ to extract the time scale $T_{\text{LZ}}$.}
\label{Fig:LZfitting}
\end{center}
\end{figure}

We study quantum annealing on random unit disk graphs, with $N$ vertices and density $\rho$. We take the limit of $\Delta_{0}, \Omega_{0} \ll U$, where the non-independent sets are pushed away by large energy penalties and can be neglected. In the experiment, this corresponds to the limit where the Rydberg interaction energy is much stronger than other energy scales. In this limit, we restrict our wavefunction to the subspace of all independent sets, i.e.
\begin{equation}
\mc{H}_{\rm IS} = \{\ket{\psi}:  n_v n_w\ket{\psi} = 0 \text{ for any } (v,w)\in E\},
\end{equation}
in our numerical simulation, which allows us to access a much bigger system size up to $N \sim 50$ since $\dim(\mc{H}_{\rm IS})\ll 2^N$. First, the subspace of all independent sets is found by a classical algorithm, the Bron-Kerbosch algorithm \cite{Bron:1973dm}, and the Hamiltonian in Eq.~\eqref{Eq:QA} is then projected into the subspace of all independent sets. The dynamics with the time-dependent Hamiltonian is simulated by dividing the total simulation time $t$ into sufficiently small discrete time $\tau$ and at each small time step, a scaling and squaring method with a truncated Taylor series approximation \cite{AlMohy:2011iw} is used to perform the time evolution without forming the full evolution operators. 

We first consider the time scale needed for adiabatic quantum annealing to work. Typically, this is governed by the minimum spectral gap, $\gap$: the runtime required is $T = O(1/\gap^2)$. However, the minimum spectral gap is ambiguous when the final ground state is highly degenerate, since it is perfectly legitimate for the state to couple to an instantaneous excited state as long as it comes down to the ground state in the end. For a generic graph, there can be many distinct maximum independent sets (the ground state of $H_{P}$ is highly degenerate). So instead of finding the minimum gap, we take a different approach to extract the adiabatic time scale.

\begin{figure}[t]
\begin{center}
\includegraphics[width=\linewidth]{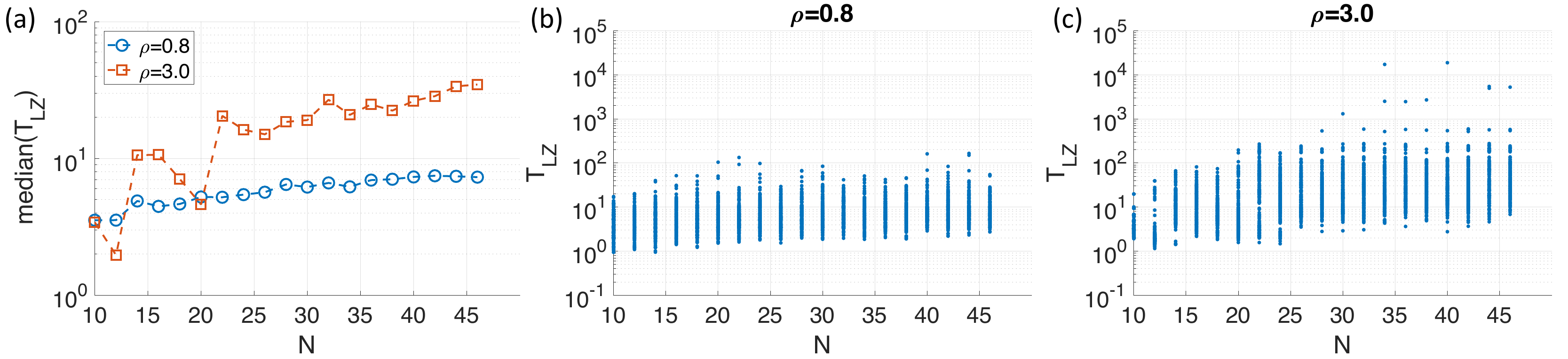} 
\caption{The adiabatic time scale $T_{\text{LZ}}$ at some fixed densities. For each system size up to $N = 46$, 200 random unit disk graphs are simulated. (a) Median $T_{\text{LZ}}$. (b) and (c): $T_{\text{LZ}}$ for individual instances for $\rho = 0.8$ and $\rho = 3$. }
\label{Fig:MIS_T_LZ_Scaling}
\end{center}
\end{figure}

In the adiabatic limit, the final ground state population (including degeneracy) takes the form of the Landau-Zener formula $P_{\text{MIS}} \approx 1 - e^{a-T/T_{\text{LZ}}}$, where $a$ is a constant and $T_{\text{LZ}}$ is the adiabatic time scale. In the nondegenerate case, one typically has $T_{\text{LZ}} = O(1/\gap^{2})$. In the more general case, we extract the time scale $T_{\text{LZ}}$ by fitting to this expression. However, the simple exponential form holds only in the adiabatic limit, where $T \gtrsim T_{\text{LZ}}$. Hence, for each graph instance, we search for the minimum $T$ such that the equation holds: we adaptively double $T$ iteratively (from $T_{\min} = 5$) until we find the minimum $T^{*}$ such that $P_{\text{MIS}} > 0.9$, at which we assume the time evolution lies in the Landau-Zener regime; we then simulate the dynamics for another three time points $1.5T^{*}$, $2T^{*}$, and $2.5T^{*}$, before finally fitting to the equation from $T^{*}$ to $2.5T^{*}$ to extract the time scale $T_{\text{LZ}}$. The fitting is remarkably good for most instances (see Fig.~\ref{Fig:LZfitting} for some examples), and we drop the few graphs where the goodness-of-fit $R^{2} < 0.99$.
We perform this procedure to extract $T_\text{LZ}$ for up to 30 graph instances at each $N$ and $\rho$, and take their median;
this produces the full phase diagram in terms of $T_{\text{LZ}}$ as plotted in Fig.~4(a) of the main text.

Here, in Fig.~\ref{Fig:MIS_T_LZ_Scaling}, we also look at the scaling of $T_{\text{LZ}}$ with $N$ at some fixed densities $\rho = 0.8$ (below the percolation threshold) and $\rho = 3$ (above the percolation threshold). We simulated quantum annealing and extracted $T_{\text{LZ}}$ for 200 random UD graphs at each $N$ up to $N = 46$. As seen in Fig.~\ref{Fig:MIS_T_LZ_Scaling}(a), we can see a clear separation between $\rho = 0.8$ and $\rho = 3$, but the scaling of $N$ is unclear, due to finite-size effect: from the performance of the classical algorithm shown in the main text, one may need to go to $N \gtrsim 100$ to see the true scaling. Fig.~\ref{Fig:MIS_T_LZ_Scaling}(b) and (c) also show the spread of $T_{\text{LZ}}$ for each instance. Note that some hard instances require significantly longer $T_{\text{LZ}}$ than the typical instance, and even on average we can see $T_{\text{LZ}} > 10$ for $\rho =3, N \gtrsim 20$. 


\begin{figure}[h]
\begin{center}
\includegraphics[width=0.5\linewidth]{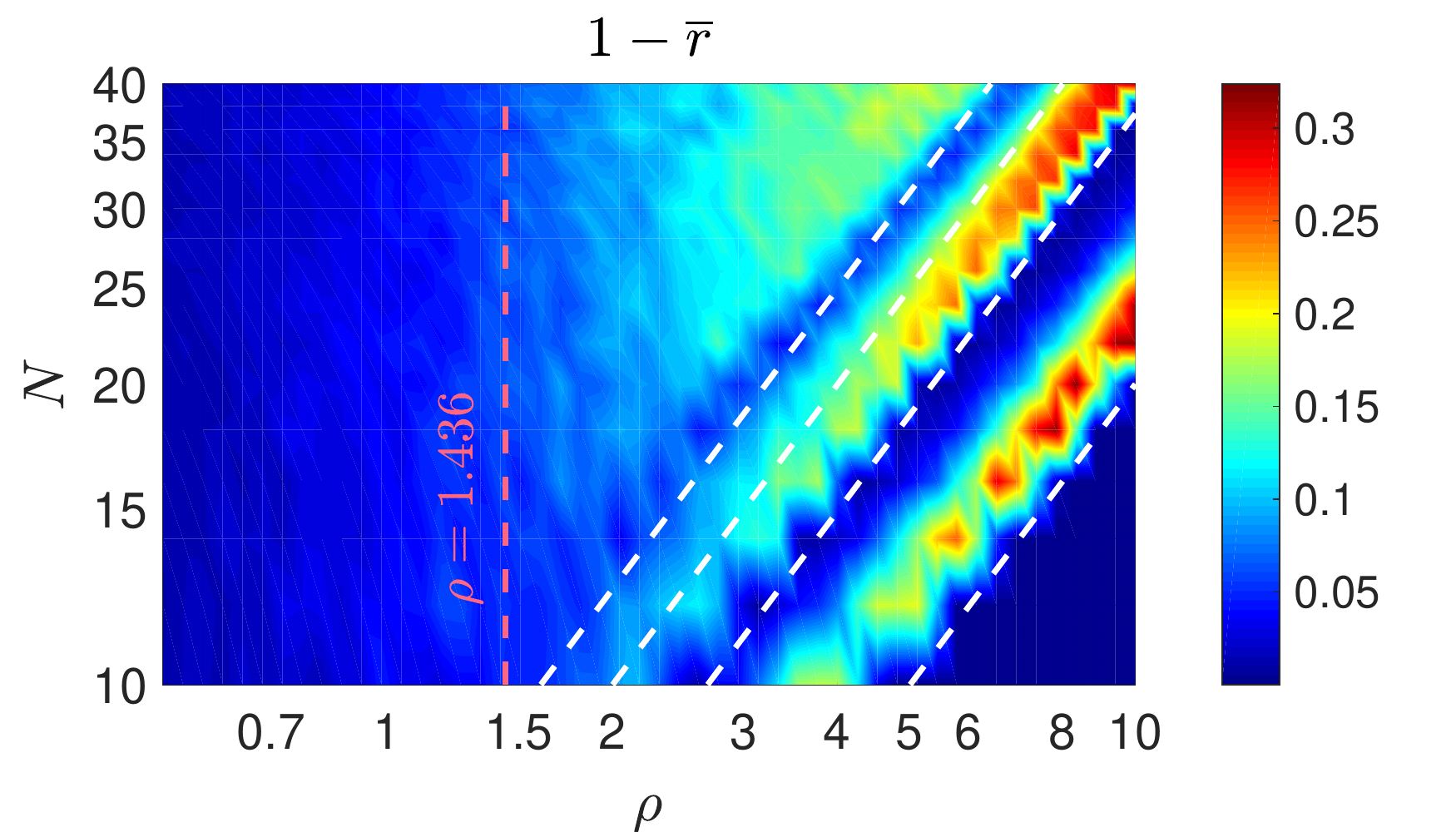} 
\caption{Phase diagram of approximation ratio $r$ for non-adiabatic quantum annealing with $T=10/\Omega_0$, averaged over 30 graph instances per $N$ and $\rho$ (the same instances as in Fig.~4(b) in the main text). Red dashed line corresponds to percolation ratio at $\rho=\rho_c\approx1.436$\cite{Mertens:2012fr}. White dashed line correspond to optimal disk packing\cite{Musin:2015dz}.}
\label{Fig:MIS_PhaseDiag_ratio}
\end{center}
\end{figure}

In the main text, we focused mainly on the capacity of the algorithms to solve MIS exactly. It is also interesting to ask whether the algorithms can solve MIS approximately, in the sense of finding an independent set as large as possible. For quantum algorithms, we use the approximation ratio $r$ to gauge their performance in terms of approximation. For a quantum algorithm (such as a quantum annealer) that outputs a state $\ket{\psi_f}$, we define $r = \sum_i\braket{\psi_f}{n_i|\psi_f}/ |\text{MIS}|$, where $|\text{MIS}|$ is the size of the MIS. In other words, $r$ quantifies the ratio of the average independent-set size found by measuring the output quantum state to the maximum independent-set size. Fig.~\ref{Fig:MIS_PhaseDiag_ratio} shows an analogous phase diagram in terms of the approximation ratio $r$ by running quantum annealing at a fixed time $T=10$. It displays qualitatively the same features as the ground state population in the main text, but the finite-size effect is stronger due to the small discrete $|\text{MIS}|$ values at large densities. 

\subsection{QAOA for MIS}

In this section, we explain how we simulate the Quantum Approximate Optimization Algorithm to solve Maximum Independent Set Problems.

\subsubsection{Quantum approximate optimization algorithm}
Suppose we are to find MIS on a given a graph $G=(V,E)$.
The $p$-level QAOA for MIS, suggested first by \cite{Farhi:2014wk}, is a variational algorithm consisting of the following steps:
\begin{itemize}
\item[(i)] Initialization of the quantum state in $\ket{\psi_0}=\ket{0}^{\otimes N}$.
\item[(ii)] Preparation of variational wavefunction 
\begin{equation}\label{QAOAv1}
\ket{\psi_p(\vgamma,\vbeta)}=\exp(-i\beta_p H_Q) \prod_{k=1}^{p-1} \exp(-i\gamma_k H_P)\exp(-i\beta_k H_{Q})\ket{\psi_0},
\end{equation}
where $H_P = \sum_{v\in V} -\Delta n_v + \sum_{(v,w) \in E} U n_v n_w$, and $H_Q=\sum_{v\in V} \Omega \sigma_v^x + \sum_{(v,w) \in E} U n_v n_w$. The parameters $\vgamma\in\mathds{R}^{p-1}$ and $\vbeta\in\mathds{R}^p$, specify the variational state.
\item[(iii)] Measurement of $H_P$.
\end{itemize}

The three steps (i)-(iii) are iterated and combined with a classical optimization of the variational parameters in order to minimize $\mean{\psi_p(\vgamma,\vbeta)|H_P|\psi_p(\vgamma,\vbeta)}$.
\subsubsection{Alternative formulation}
We are interested in the $U\gg |\Omega|,|\Delta|$ limit, where the variational search is restricted to the subspace $\mc{H}_{\rm IS}$ spanned by independent sets, such that the algorithm does not need to explore states that can be directly excluded as MIS candidates.
In this limit, we can write
\begin{align}
H_Q=\sum_v \mc{P}_{\rm IS}\Omega\sigma_x\mc{P}_{\rm IS}, \qquad H_P= \sum_{v\in V} -\Delta n_v,
\end{align}
where $\mc{P}_{\rm IS}$ is a projector onto the independent set subspace $\mc{H}_{\rm IS}$. 
Evolution with $H_P$ thus reduces to simple rotation of individual spins around the $z$ axis. Since
\begin{align}
\exp(-i\gamma H_P)\exp(-i\beta H_Q)=\exp\lr{-i\beta \Omega \sum_v \mc{P}_{\rm IS} \lr{\ket{0}_v\bra{1}e^{i\gamma}+\rm{h.c.}}\mc{P}_{\rm IS}}\exp(-i\gamma H_P),
\end{align}
we can commute all the unitaries generated by $H_P$ in \eqref{QAOAv1} to the rightmost side until they act (trivially) on the initial state. Thus, we can rewrite the state $\ket{\psi_p(\vbeta,\vgamma)}$ as
\begin{equation}\label{QAOAv2}
\ket{\psi_p(\vec{\gamma},\vec{\beta})}=\prod_{k=1}^p\exp\lr{-it_k \Omega \sum_v \mc{P}_{\rm IS} \lr{\ket{0}_v\bra{1}e^{i\phi_k}+\rm{h.c.}}\mc{P}_{\rm IS}}\ket{\psi_0},
\end{equation}
where we identify
\begin{align}
\phi_k=\sum_{j\ge k}\gamma_j, \qquad t_k=\beta_k
\end{align}
Thus we recover the formulation of QAOA given in the main text, which is equivalent to \eqref{QAOAv2} for $U\gg\Omega$.

\subsubsection{Numerical simulations: preliminaries}
In our numerical study, we work in the formulation of QAOA corresponding to \eqref{QAOAv1} and take $\Delta=\Omega=1$.
Again, we work in the limit where $U\gg 1$ so that we can restrict our Hilbert space to the independent set subspace $\mc{H}_{\rm IS}$;
this allows us to efficiently simulate system sizes up to $N \sim 50$.
We prepare the state as in Eq.~\eqref{QAOAv1}, and then measure the expectation value of $H_P$, which is the objective function that we seek to minimize:
\begin{equation}
F_p(\vec{\gamma},\vec{\beta}) = \bra{\psi_p(\vec{\gamma},\vec{\beta})}H_P\ket{\psi_p(\vec{\gamma},\vec{\beta})}.
\end{equation}
This sequence of state-preparation and measurement of objective function is then fed as a subroutine to a classical optimization algorithm, which is used to find good parameters $(\vec\gamma,\vec\beta)$ with the lowest possible $F_p$.

\vspace{5pt}
\noindent
\textbf{Classical Optimization Algorithms}---
Generally, classical optimization algorithms work by starting with some initial point in the QAOA parameter space $(\vec\gamma,\vec\beta)$, and iteratively find new points $(\vec{\gamma}', \vec\beta')$ using information from the current point $(\vec\gamma, \vec\beta)$, with the hope that a new point produces a lower value of the objective function $F_p(\vec\gamma',\vec\beta') \le F_p(\vec\gamma,\vec\beta)$.
We first describe some free parameters and stopping criteria that apply to these classical optimization algorithms:
\begin{itemize}
\item $\delta$ -- step tolerance, If the optimization algorithm attempts to go to a new set of parameters $(\vec{\gamma}', \vec\beta')$ such that $|\vec\gamma'-\vec\gamma|^2 + |\vec\beta'-\vec\beta|^2 \le \delta^2$, then the algorithm terminates and outputs the smallest value of $F_p$ seen so far. When the algorithm calls for a numerical computation of the gradient using the finite-difference method, we also take $\delta$ as our increment size, e.g. $\partial F_p/\partial \gamma_i \simeq [F_p(\gamma_i+\delta)-F_p(\gamma_i)]/\delta$.
\item $\epsilon$ -- objective function tolerance. If the optimization algorithm finds that the change in the value of objective function is smaller than $\epsilon$, $|F_p(\vec\gamma',\vec\beta')-F_p(\vec\gamma,\vec\beta)|\le \epsilon$, then the algorithm terminates and outputs the smallest value of $F_p$ seen so far.
\item $\epsilon_M$ -- measurement precision. This is the target precision to which we compute the objective function $F_p$.
When simulating QAOA with measurement projection noise, this parameter determines the number of measurements necessary to obtain a good averaged value of $\mean{H_P}$.
\end{itemize}

In our numerical study, we consider two classical optimization algorithms: the BFGS quasi-Newton algorithm\cite{BFGS1,*BFGS2,*BFGS3,*BFGS4} and the Nelder-Mead simplex algorithm \cite{Nelder:1965in}.
Specifically, the BFGS algorithm computes the gradient of the objective function at the current point $(\vec\gamma, \vec\beta)$, and uses this information to build a quadratic model of objective function and determine the approximate location of a local minimum.
This algorithm terminates the optimization routine if either step tolerance or objective function tolerance is reached.
However, computing the gradient in high-dimensional parameter space can be inefficient, especially when the cost of measurement is being considered.
Hence, we also consider the Nelder-Mead simplex algorithm that does not involve gradients; this algorithm terminates the optimization routine when \emph{both} step tolerance and objective function tolerance are reached.
These algorithms are implemented in the standard library of MATLAB R2017b via \texttt{fminunc} and \texttt{fminsearch} functions, respectively.

\vspace{5pt}
\noindent
\textbf{Heuristic ansatz for optimizing QAOA at deep depths}---
For high levels of QAOA with deep circuit depths, optimization can be difficult because of large dimension of QAOA parameter space.
Due to the non-convex nature of the objective function $F_p$, one often need to optimize starting from many points to have a better chance of finding the global minimum in the QAOA parameter space.
Typically, a brute-force approach starting from random points will require $2^{O(p)}$ initial points to find the global minimum.
Nevertheless, as discussed in more detail in an upcoming study \cite{MaxCutpaper}, we discover patterns in the optimal QAOA parameters for most instances of the MIS problem.
Based on these patterns, we develop a heuristic strategy for choosing good initial points when optimizing QAOA parameters at intermediate level $p$ \cite{MaxCutpaper}.
We find that this strategy allows us to find quasi-optimal parameters that's often nearly as good as the true global minimum, but in time only $O(\poly(p))$.

We now describe our heuristic strategy optimizing QAOA parameters when applied to MIS problems.
We start with level $p=3$ and optimize from an educated guess of initial point $(\gamma_1,\gamma_2,\beta_1,\beta_2,\beta_3) \approx (1.73,-1.77,0.19,1.02,0.39)$ based on the averaged optimal QAOA parameters from 20 instances.
When the optimization algorithm terminates with the optimized parameters $(\vgamma_{(p)}^L, \vbeta_{(p)}^L)$ for level $p$, we move on to level $p+1$ with the initial point $(\vgamma_{(p+1)}^0, \vbeta_{(p+1)}^0)$ obtained by linear interpolation:
\begin{align}
\left[\vgamma^{0}_{(p+1)}\right]_1 & = \left[\vgamma^{L}_{(p)}\right]_1, \quad \left[\vgamma^{0}_{(p+1)}\right]_{p}  = \left[\vgamma^{L}_{(p)}\right]_{p-1} , 
\quad
\left[\vgamma^{0}_{(p+1)}\right]_i  = \tfrac{i-1}{p-1} \left[\vgamma^{L}_{(p)}\right]_{i-1}+ \tfrac{p-i}{p-1}\left[\vgamma^{L}_{(p)}\right]_i,  \\
\left[\vbeta^{0}_{(p+1)}\right]_1 & = \left[\vbeta^{L}_{(p)}\right]_1, \quad \left[\vbeta^{0}_{(p+1)}\right]_{p+1}  = \left[\vbeta^{L}_{(p)}\right]_{p} , 
\quad
\left[\vbeta^{0}_{(p+1)}\right]_j  = \tfrac{j-1}{p} \left[\vbeta^{L}_{(p)}\right]_{j-1}+ \tfrac{p-j+1}{p}\left[\vbeta^{L}_{(p)}\right]_j, \\
 & \text{when } \quad 2\le i \le p-1 \text{ and } 2\le j \le p.\nonumber
\end{align}
Here, we denote $[\vgamma]_i\equiv \gamma_i$ as the $i$-element of parameter vector $\vgamma$.
This strategy takes advantage of the observation that there is often a set of optimal QAOA parameters that change smoothly from level $p$ to $p+1$.
Using this strategy, we have been able find good parameters for QAOA at levels as large as $p=50$.

\subsubsection{Simulation neglecting measurement cost}
In order to understand the potential of QAOA by finding the best possible optimal parameters, our initial numerical study neglects the cost of measurements and lets the simulated quantum processor output the exact values of $F_p(\vec\gamma,\vec\beta)$.
In other words, we effectively choose $\epsilon_M=0$.
In addition, we calculate the gradient of the objective function $\nabla F_p = (\partial F_p/\partial\vec\gamma, \partial F_p/\partial\vec\beta)$ analytically instead of using the finite-difference method.
The BFGS quasi-Newton algorithm is used in these simulations to find local minima in the QAOA parameter space, where we set the tolerances at $\epsilon=\delta=10^{-6}$.
This approach of ignoring measurement cost allows us to find optimal QAOA parameters more efficiently.
Using our heuristic ansatz mentioned above, we find optimal QAOA parameters up to $p=50$.
Our results show that optimized QAOA can achieve better performance than simple quantum annealing schedules, as shown for the instance in Fig.~4(c) of the main text.
Nevertheless, this approach does not represent a realistic simulation of actual QAOA implementations, where measurements are not free and $\epsilon_M$ is necessarily finite.

\subsubsection{Simulated experiment with measurement projection noise}
In actual experiments, the value of the objective function $F_p$ can only be determined approximately by averaging over many measurements, each projecting the wavefunction onto a possible outcome.
In our numerical simulation, we account for this effect by performing full Monte Carlo simulation of actual measurements, where the quantum processor outputs only an approximate value of the objective function obtained by averaging over $M$ measurements:
\begin{equation}
\tilde{F}_p = \frac{1}{M}\sum_{i=1}^M f_{p,i}, \quad
f_{p,i} \text{ is a random variable where }
\Pr(f_{p,i} = -k) = \braket{\psi_p(\vec\gamma,\vec\beta)}{\Pi_k|\psi_p(\vec\gamma,\vec\beta)},
\end{equation}
and $\Pi_k$ is the projector onto subspace where $H_P = -k$, spanned by independent sets of size $k$.
Note that when $M\to \infty$, we obtain $\tilde{F}_p \to F_p=\braket{\psi_p(\vec\gamma,\vec\beta)}{H_P|\psi_p(\vec\gamma,\vec\beta)}$ with perfect precision.
In order to achieve finite precision $|\tilde{F}_p - F_p | \sim \epsilon_M$,
we accumulate measurements until the standard error of the mean falls below the precision level.
In other words, for each evaluation of $F_p(\vgamma,\vbeta)$, the number of measurements $M$ we perform is set by the following criterion:
\begin{equation}
\sqrt{\frac{1}{M(M-1)}\sum_{i=1}^M(f_{p,i}-\bar{F}_{p,M})^2} \le \epsilon_M, 
\quad\text{where}\quad
\bar{F}_{p,M} = \frac{1}{M}\sum_{i=1}^M f_{p,i}.
\end{equation}
Roughly speaking, $M\approx \Var(\hat{F}_p)/\epsilon_M^2$.
To mitigate finite-sample-size effects, we also require at least 10 measurements ($M\ge 10$) be performed for each evaluation of $F_p$.

Using this approach, we simulate experiments of optimizing QAOA with measurement projection noise.
The above-mentioned heuristic ansatz is utilized, as we start with an educated guess of initial QAOA parameters at $p=3$ and optimize until tolerance levels are reached.
To illustrate the usefulness of our heuristic ansatz, we also simulate starting with random choices of initial parameters for optimization, and compare their performance.
In these simulations, the approximate value of $\tilde{F}_p(\vgamma,\vbeta)$ is returned whenever the classical optimization algorithm requests an evaluation of $F_p(\vgamma,\vbeta)$ from the simulated quantum processor.
This includes, for example, when the BFGS algorithm numerically computes gradients to find optimal parameters by the finite-difference method: $\partial F_p/\partial \gamma_i \approx [\tilde{F}_p(\gamma_i+\delta) - \tilde{F}_p(\gamma_i)]/\delta$.
The history of measurements is stored throughout the entire simulated experiment, which allows us to keep track of the largest independent set $\text{IS}(m)$ found after $m$-th measurement.
We repeat this numerically simulated experiment many times with different pseudo-random number generation seeds, and average over their histories.
In Fig.~4(d) of the main text, we show an example instance where the simulated experiments are run with $\epsilon=\delta=0.2$, and $\epsilon_M = 0.05$, with and without our heuristic ansatz.

\section{Generalization to arbitrary graph structure}
\subsection{Stroboscopic evolution}
As mentioned in the main text, one can generalize our implementation to address MIS problems on graphs, $G=(V,E)$ that are beyond the UD paradigm. 

To do so, let us first note that all quantum algorithms we discussed in the main text require evolution with a Hamiltonian $H(t)=\sum_{v}\Omega_v(t)\sigma_v^x-\Delta_v(t)n_v+\sum_{(u,v)\in E} U n_u n_v$.
In particular, we are interested in the situation where $U\gg |\Omega|,|\Delta|$, such that the dynamics is effectively restricted to the independent set space $\mc{H}_{\rm IS}$. To generate such evolution with a Hamiltonian corresponding to a general graph structure, let us consider a Trotterized version of the time evolution operator 
\begin{align}
\mc{T}\exp{ \lr{-i\int_0^T dt\,H(t)}}\simeq \prod_{j}\mc{U}(t_j)\equiv\prod_{j}\exp{ \lr{-i(t_{j+1}-t_{j})H(t_j)}},
\end{align}
where we have sliced the time interval $[0,T]$ defining times $t_j$ such that $\sum_j t_j=T$ and $t_{j+1}-t_j\ll \sqrt{D_{\rm max}}\Omega(t_j),|\Delta(t_j)|$. Here $D_{\rm max}$ denotes the maximum degree of the graph. We further Trotterize each $\mc{U}(t_j)$ as follows
\begin{align}
\mc{U}(t_j)\simeq \prod_{v=1}^N\mc{U}_v(t_j)\equiv\prod_{v=1}^N \exp\lr{-i(t_{j+1}-t_j)\lr{\Omega_v(t_j)\sigma_v^x-\Delta_v(t_j)n_v+\frac{1}{2}\sum_{u\in \mc{N}(v)} U n_u n_v)}},
\end{align}
that is we split it into a product of terms $\mc{U}_v$ that each are associated with the evolution of one spin, $v$. Here $\mc{N}(v)$ denotes the neighbors of $v$ on the graph. Note that in the $U\rightarrow\infty$ limit we are interested this can be written as 
\begin{align}\label{Trotterstep}
 \mc{U}_v(t_j)=\exp\lr{-i(t_{j+1}-t_j)\lr{\Omega_v(t_j)\sigma_v^x-\Delta_v(t_j)n_v}}\prod_{u\in \mc{N}(v)}\ket{0}_u\bra{0}.
 \end{align}
This is a simple single qubit rotation of atom $v$, condition on the state of the atoms corresponding to neighbors of $v$ being $\ket{0}$. If at least one of the neighbors is in state $\ket{1}$, atom $v$ does not evolve. 

\subsection{Implementation using qubit hyperfine encoding}
One approach to realize the corresponding dynamics with individually controlled neutral atoms can be designed as follows.  We consider an implementation where qubits states $\ket{0}$ and $\ket{1}$ are encoded in two (non-interacting) hyperfine states, in the internal atomic ground state manifold. We position all atoms on the points of a 2D square lattice with distance $g$. To realize a single step, $\mc{U}_v(t_j)$, we first excite all atoms, $u$, that correspond to neighbors of $v$ on the graph ($u\in \mc{N}(v)$), selectively from the state $\ket{1}$ to a Rydberg $S$-state $\ket{1'}$. We choose a grid length $g\ll r_B$ such that none of the atoms $u\in \mc{N}(v)$ interact during this process. Then we drive atom $v$, to realize the single qubit rotation in the hyperfine manifold, i.e a unitary corresponding to an evolution with $\Omega_v(t_j)\sigma_v^x-\Delta_v(t_j)n_v$, where ${\sigma}_v^x$ couples the two hyperfine qubits states of atom $v$, and $n_v=\ket{1}_v\bra{1}$ counts if atom $v$ is in hyperfine state $\ket{1}$. To realize this rotation this we use an individually addressed two-step excitation that couples the two hyperfine states $\ket{0}$ and $\ket{1}$ of atom $v$ via a transition through a Rydberg $P$-state. If all atoms $u$ are in the state $\ket{0}$, then this process is not disturbed, but if at least one of the neighbors is the Rydberg $S$ state, the strong $S-P$ interaction gives rise to a blockade mechanism that prevents the rotation of the qubit $v$, thus realizing exactly \eqref{Trotterstep}. Note this requires a different scale of interaction length of the blockade radius for two atoms in the Rydberg $S$-states on one hand, and the $S$-$P$ blockade radius on the other hand. This can be readily realized by noting  that  these two interactions scale differently with separation of the atoms: the $S$-$P$ interactions decay as $1/x^3$, i.e. much slower than the $S$-$S$ interactions that scale like $1/x^6$ [refs, thompson, nature], which should allow one to implement collective gate with high fidelity. 
\bibliography{HannesBib,additional_refs}